\documentclass[11pt,a4paper]{article}
\pdfoutput=1 
\usepackage{amsmath}
\usepackage{amsthm,amssymb}

\usepackage{graphicx}

\numberwithin{equation}{section}

\newcommand{\be}{\begin{equation}}
\newcommand{\ee}{\end{equation}}
\newcommand{\ben}{\begin{equation*}}
\newcommand{\een}{\end{equation*}}
\newcommand{\bea}{\begin{eqnarray}}
\newcommand{\eea}{\end{eqnarray}}
\newcommand{\bean}{\begin{eqnarray*}}
\newcommand{\eean}{\end{eqnarray*}}

\newcommand{\bi}{\begin{itemize}}
\newcommand{\ei}{\end{itemize}}

\newcommand{\bs}{\left [ \begin{matrix}}
\newcommand{\fs}{\end{matrix} \right ] }

\newcommand{\bm}{\left ( \begin{matrix}}
\newcommand{\fm}{\end{matrix} \right ) }

\begin{document}

\title{Forcing Adsorption of a Tethered Polymer by Pulling}
\author{J Osborn$^1$ and T Prellberg$^2$ \vspace{1mm}\\
  \footnotesize
  \begin{minipage}{6cm}
    $^1$ Mathematical Sciences Institute,\\
    Australian National University, \\Canberra, ACT 0200, Australia.\\
    \texttt{Judy-anne.Osborn@anu.edu.au}
   \end{minipage}
   \begin{minipage}{6cm}
$^2$ School of Mathematical Sciences\\
Queen Mary University of London\\
Mile End Road, London E1 4NS, UK\\
\texttt{t.prellberg@qmul.ac.uk}
\end{minipage}
}

\maketitle  

\begin{abstract}
We present an analysis of a partially directed walk model of a polymer
which at one end is tethered to a sticky surface and at the other end 
is subjected to a pulling force at fixed angle away from the point of 
tethering. Using the kernel method, we derive the full generating function 
for this model in two and three dimensions and obtain the respective phase diagrams.

We observe adsorbed and desorbed phases with a thermodynamic phase transition
in between. In the absence of a pulling force this model has a second-order
thermal desorption transition which merely gets shifted by the presence of a
lateral pulling force. On the other hand, if the pulling force contains a 
non-zero vertical component this transition becomes first-order.

Strikingly, we find that if the angle between the pulling force and the 
surface is beneath a critical value, a sufficiently strong force will induce
polymer adsorption, no matter how large the temperature of the system.

Our findings are similar in two and three dimensions, an additional feature
in three dimensions being the occurrence of a reentrance transition at 
constant pulling force for small temperature, which has been observed
previously for this model in the presence of pure vertical pulling. 
Interestingly, the reentrance phenomenon vanishes under certain pulling angles,  with details depending on how the three-dimensional polymer is modeled.

\end{abstract}
 
\section{Introduction}
There is ongoing interest in the study of linear polymers and their conformal entropy using directed and partially directed walk models, both from theoretical \cite{Rensburg2000, Rensburg2003} and experimental \cite{kumar2010} perspectives.  These models are of particular interest because it is often possible to obtain exact solutions for their generating functions, partition functions and free energy.  Thus a precise understanding of the phase transitions that the models can exhibit may be obtained, which can then be used to guide and interpret laboratory experiments on physical polymers \cite{kumar2010}.  

In appropriate physical settings, for example when the polymer undergoes surface adsorption, good qualitative agreement has been shown between the phase behavior of directed walk models and that of  self avoiding walk (SAW) models \cite{mishra2003}, the latter of which exhibit more of the configurational possibilities available to real polymers, but are less tractable as mathematical models.    Under the presence of a pulling force, this is intuitive, because the application of such a force to a physical polymer tends to straighten it, making it behave more like a directed or partially directed path \cite{mishra2005}. 

In this paper we consider a two-dimensional partially directed walk model for a polymer which at one end is tethered to a sticky surface and at the other end is subjected to a pulling force in a fixed direction away from the point of tethering.  We then extend this model into three dimensions in two different ways. The first extension is `minimally three-dimensional', in that its projection onto the surface is fully directed, and as a consequence walks adsorbed in the surface are fully directed.  This extension has previously been considered in \cite{orlandini2004} for a purely vertical pulling force. The second three-dimensional extension we consider is different from the first in that walks adsorbed in the surface are partially directed.  To our knowledge, this model has not been considered previously.

Another novelty in our model is in the variable angle with which the pulling force may be applied to our tethered polymers.  In similar previously published literature the pulling force was restricted to be vertical \cite{mishra2005, orlandini2004, owczarek2009, owczarek2010, iliev2010, bhattacharya2009, krawczyk2004}, a limitation which does not apply to the technology of optical tweezers which is used by experimentalists in `pulling-force' experiments carried out on real polymers. Changing the pulling direction has been considered previously in anisotropic self-interacting
models \cite{kumar2007,rajesh2008,jensen2010}.

Our use of partially directed walks, rather than simpler  directed models such as Dyck, Ballot and Motzkin paths, is for the purpose of  allowing the extra degree of freedom required to enable polymers to extend horizontally when pulled upon with a force containing a horizontal component.

\section{Partially Directed Walks}

Let $\mathcal{U} = \mathbb{Z} \times \mathbb{Z}_{\ge 0}$ be the square integer lattice in the upper half plane.  Let $p=v_0e_0v_1e_1 ... e_Lv_L$ be a path of length $L$ consisting of an alternating sequence of vertices and edges such that each vertex $v_i \in \mathcal{U}$ and each edge $e_i:=(v_{i-1},v_{i})$ is of one of the following three types:
\bea
v_i - v_{i-1} = 
\begin{cases}
(0,1) & \text{a vertical step \emph{up}} \\
(1,0) & \text{a horizontal step \emph{across}} \\
(0, -1) & \text{a vertical step \emph{down}}
\end{cases}
\eea

We impose three additional constraints on walks in our model.  
The first is tethering to the origin, imposed by the condition
$
v_0=(0,0).
$
The second is that paths are self-avoiding, that is, an up step may not be immediately followed by a down step, or vice versa.  We call such paths tethered \emph{partially directed walks}.  
The third constraint is that paths must always start and end on a horizontal step.
This latter constraint makes negligible difference to the physics, and is introduced for computational convenience in finding the generating function.  

\begin{figure}[htbp]
\begin{center}
\includegraphics[width=0.8\textwidth]{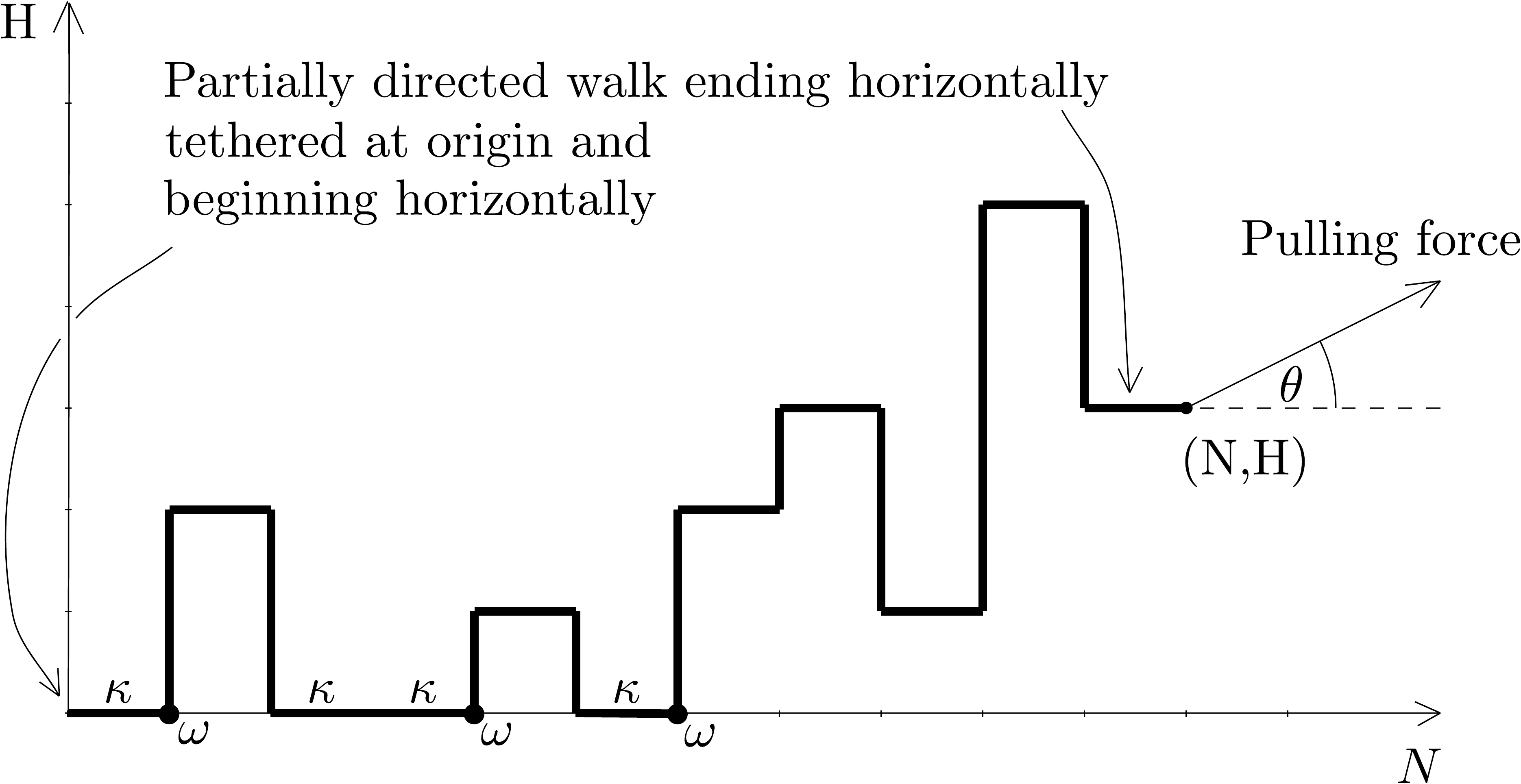}  
\caption{A partially directed walk with weight $x^{11}y^{17}\kappa^{4}\omega^{3}\mu^{3}$.  For clarity in the diagram, the weights on the edges leaving the surface are indicated visually by dots on their initial vertices.}  
\label{fig:walk2Dc}
\end{center}
\end{figure}

Geometrically relevant parameters of a walk are:
\begin{subequations}
\bea
N &:=& \text{the number of \emph{horizontal} edges in the walk,}\\
M &:=& \text{the total number of \emph{vertical} edges in the walk,}\\
K &:=& \text{the number of \emph{horizontal} edges of the walk lying in the surface,}\\
R &:=& \text{the number of \emph{vertical} edges of the walk leaving the surface,}\\
H &:=& \text{the final \emph{height} of the walk.}
\eea
\end{subequations}
We define the \emph{weight} of a partially directed walk to be
\be \label{eq:C}
x^N y^M  \kappa^K \omega^R \mu^H.
\ee
An example of a partially directed walk and its weight is illustrated in Figure~\ref{fig:walk2Dc}.  Also illustrated is a pulling-force at angle $\theta$ away from the $N$-axis.

Note that the conjugate pair of variables $\omega$ and $R$, which keeps track of vertical departures from the surface, is introduced for later convenience in our consideration of our second three dimensional model.  For the purposes of the two-dimensional and first three-dimensional model, these variables are unnecessary, and setting $\omega=1$ in the following calculations simplifies them considerably.

Also note that setting $\omega=\kappa$ corresponds to considering a model in which vertices of the walk that are in the surface are weighted. We have considered both edge and vertex weighted models, and have found negligible differences.
For simplicity we only present results for the edge-weighted models in this paper. Results for vertex-weighted models
follow \emph{mutatis mutandis}. 

\section{Exact Solution of the Generating Function}

We define a generating function
\begin{equation}
G(x, y, \kappa, \omega; \mu) = \sum_{N,M,K,R,H \ge 0}c_{N,M,K,R,H}x^Ny^M \kappa^{K}\omega^R \mu^H  
\end{equation}
in terms of the path weight function given in Equation~(\ref{eq:C}), where $c_{N,M,K,R,H}$ counts the number of
different configurations with given parameters $N$, $M$, $K$, $R$, and $H$.
A functional relation for the generating function may be obtained by considering paths ending at a fixed height.
Suppressing the first four variables, we abbreviate $G(\mu):=G(x,y,\kappa, \omega;\mu)$.  A combinatorial decomposition of the set of all walks with respect to generation by addition of `hooks' of arbitrary height and ending in a horizontal step leads to the following functional equation.
\begin{subequations}
\begin{align}
G(\mu) =&\; \kappa x &\text{ horizontal step at height 0} \label{eq:kappax}\\
&+ G(\mu) \Bigg (x &\text{horizontal step at height $>0$} \label{eq:Ghoriz}\\
&+ \frac{ y \mu }{1 - y \mu}\,x
&\text{vertical steps up, then horizontal}\\
&+ \frac{y/\mu}{1 - y/\mu}\,x\Bigg )
&\text{ vertical steps down, then horizontal} \label{eq:Gdown}\\
&
- G(y) \frac{y/\mu}{1-y/\mu}\,x
&\text{removes walks descending below the surface} \label{eq:overCount}
\\
&
+ G(y) (\kappa-1) \,x  \label{eq:kappaxG}
&\text{ plus contact weights for edges in the surface}\\
&
+ G(0) (\omega-1) \frac{ y \mu }{1 - y \mu} \,x  \label{eq:omega}
&\text{plus contact weights for edges leaving the surface}
\end{align}
\end{subequations}

The derivation of the functional equation is summarized in Lines~(\ref{eq:kappax})--(\ref{eq:omega}).   Since our paths are required to start and end in a horizontal step, the shortest possible path consists of a single horizontal step, and is accounted for by Line~(\ref{eq:kappax}).  Lines~(\ref{eq:Ghoriz})--(\ref{eq:Gdown}) account for longer paths created by concatenating paths ending at height $H$, as encoded by the factor of $G(\mu)$, with paths containing a single horizontal step and any number of vertical steps.  
Line~(\ref{eq:Gdown}) entails some over-counting, as it includes paths  which contain vertices below the upper half plane, $\mathcal{U}$.  This over-counting is compensated for by Line~(\ref{eq:overCount}), which removes from the count any paths containing edges strictly below the $N$-axis.  

So far paths which touch but don't drop below the surface have been included without a weight $\kappa$ on those edges in the path that lie in the surface.  
Line~(\ref{eq:kappaxG}) corrects for this.  Similarly, Line~(\ref{eq:omega}) corrects for the missing weight $\omega$ on vertical edges whose initial vertex lies in the surface. 

The functional equation thus obtained is amenable to the so-called \emph{kernel method} \cite{prodinger2004}.
First we simplify the functional equation somewhat by observing that $G(0)$ and $G(y)$ are simply related by
\be \label{eq:G0y}
G(0) = \kappa x (1+G(y)),
\ee
due to the fact that any walk ending with a horizontal step in the surface must either be a single step walk or have been obtained by dropping $H$ steps down from height $H$ and adding a horizontal step in the surface. 

Using Equation~\ref{eq:G0y}, the functional equation is expressed in Equation~(\ref{eq:GmuGy}) below in a form suitable for solution by means of the kernel method,
\begin{multline} \label{eq:GmuGy}
\overbrace{\left (1-\frac{x(1-y^2)}{1-y(\mu+\frac{1}{\mu})+y^2}\right)}^{\text{kernel}}G(\mu)
=\\
\kappa x
\left (
1+(\omega-1)x\frac{y\mu}{1-y\mu}
\right ) 
-x\left ( \frac{1}{1-y/\mu}-\kappa -(\omega-1)\kappa x \frac{y\mu}{1-y\mu}\right )G(y)\;.
\end{multline}

The coefficient of $G(\mu)$ is called the \emph{kernel}.  Setting the kernel equal to zero gives a quadratic equation for $\mu$:
\be \label{eq:quadraticForMu}
y \mu^2 -(1-x+y^2+x y^2)\mu + y = 0\;.
\ee

Of the two roots, one is physically meaningful - we call it $\mu_p$  and denote the non-physical root by $\mu_{\text{np}}$.  The correct root may be identified by expanding the power series and choosing the one with correct asymptotic behavior for $G$  as $y \rightarrow 0$.  
Equivalently, the correct choice may be identified as the one which is consistent with the combinatorial interpretation of $G(y)$. 
Using this criterion we choose the root whose series expansion contains no negative exponents, and has only non-negative coefficients in that expansion.  Choosing the correct root $\mu=\mu_{\text{p}}$ gives
\be
G(y) = \frac{\kappa
\left (
1+(\omega-1)x\dfrac{y\mu_p}{1-y\mu_p}
\right ) }{\left ( \dfrac{1}{1-y/\mu_p}-\kappa -(\omega-1)\kappa x \dfrac{y\mu_p}{1-y\mu_p}\right )}\;.
\ee
Thence, back-substituting $G(y)$ into Equation~(\ref{eq:GmuGy}) gives the full general solution for $G(\mu)$; explicitly:

\begin{multline} \label{eq:Gfull}
G(x,y,\kappa, \omega;\mu)=\\ \kappa x
\frac{
\left (
1+(\omega-1)x\dfrac{y\mu}{1-y\mu}
\right ) 
- \left (
1+(\omega-1)x\dfrac{y\mu_p}{1-y\mu_p}
\right )\dfrac{\left ( \dfrac{1}{1-y/\mu}-\kappa -(\omega-1)\kappa x \dfrac{y\mu}{1-y\mu}\right )
 }{\left ( \dfrac{1}{1-y/\mu_p}-\kappa -(\omega-1)\kappa x \dfrac{y\mu_p}{1-y\mu_p}\right )}}
{1-\dfrac{x(1-y^2)}{1-y(\mu+\frac{1}{\mu})+y^2}}
,
\end{multline}
where
\be \label{eq:muPhysical}
\mu_{\text{p}} = \frac{(1 - x + y^2 + x y^2) - \sqrt{-4 y^2 + (1 - x + y^2 + x y^2)^2}}{
 2 y}\;.
\ee

\section{The Path-Length Generating Functions}

In this section we introduce the changes to the generating function (\ref{eq:Gfull}) that are needed for
our three different models. We aim to write the generating function such that we can start to analyze the
finite-step partition functions $Z_L$, for walks of length $L$, which are coefficients of $t^L$ in an expansion
$G= \sum_{L \ge 0} t^L Z_L$, where $t$ is a new variable that is conjugate to the path length $L$.

\subsection{The 2D model}

The finite-step partition function $Z_L$, for walks of length $L$ in two dimensions, is the coefficient of $t^L$ in the following expansion of the path-length generating function
\be \label{eq:G2d}
G(x=\lambda t, y=t, \kappa, \omega=1; \mu) = \sum_{L \ge 0} t^L Z_L (\lambda, \mu, \kappa)
\ee
under the substitutions 
\be \label{eq:2Dsubs}
\left (x,y,\kappa, \omega; \mu \right ) \gets \left (\lambda t, t, \kappa, 1; \mu \right ),
\ee
so that $t$ is conjugate to the \emph{path length}, $\lambda$ is conjugate to the \emph{horizontal position} and  $\mu$ is conjugate to the \emph{vertical position}.  In this way we only give a contact weight to edges in the surface.  In Section~\ref{sec:2dSingularity} we analyze the singularities of Equation~(\ref{eq:G2d})  to determine the asymptotic growth of $Z_L$ in two dimensions.

An alternative model would be obtained by weighting all vertices in the surface, leading to the consideration of $G(\lambda t, t, \kappa, \kappa; \mu)$.  We have completed both analyses and have found no significant difference, hence we only present the analysis of the case defined by Equation~(\ref{eq:G2d}).

\subsection{The first 3D model}
\begin{figure}[htbp]
\begin{center}
\includegraphics[width=0.8\textwidth]{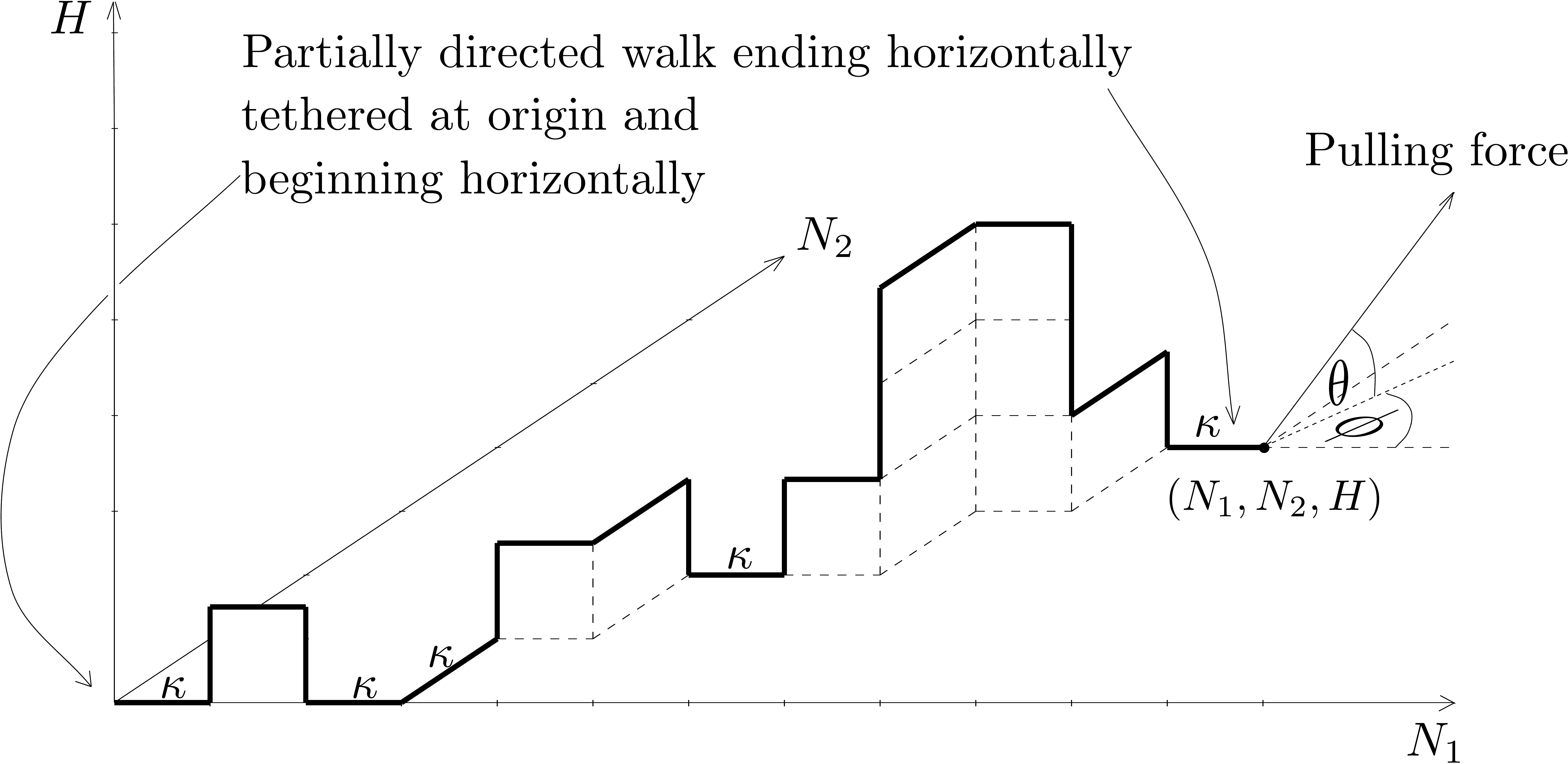}
\caption{A tethered, partially directed walk in 3 dimensions, on a sticky surface with contact weights $\kappa$ within the $N_1-N_2$ surface and a pulling-force applied at fixed angle $\phi$ away from the $N_1$-axis and angle $\theta$ away from the $N_1N_2$ surface.  This walk ends at position $(N_1, N_2, H) = (8,4,0)$ and has weight $\lambda_1^8 \lambda_2^4\, t^{21} \kappa^5 \mu^0$.}
\label{fig:walk3Db}
\end{center}
\end{figure} 

Our first three-dimensional model is obtained by a substitution which replaces any step in the $N$-direction with a step in either the $N_1$ or $N_2$ direction.  The projection of the resulting walk onto the $N_1-N_2$ plane is fully directed. This model is equivalent to the bi-colored walk model considered in \cite{iliev2010}. This corresponds to a substitution in the generating function variables given by
\be\label{eq:subst-3d1}
(x, y, \kappa, \omega; \mu) \gets \left( \lambda_1 t + \lambda_2 t,\; t, \; \kappa, \; 1; \; \mu \right)
\ee
so that $t$ is conjugate to path length, $\lambda_1$ is conjugate to  the $N_1$-coordinate, $\lambda_2$ is conjugate to the $N_2$-coordinate and $\mu$ is conjugate to the $H$-coordinate.  A three-dimensional partially directed walk of our first kind is illustrated in Figure~\ref{fig:walk3Db}; and the asymptotic growth of $Z_L$ in three dimensions is analyzed in Section~\ref{sec:3d} below.

\subsection{The second 3D model}

Our second three-dimensional model is obtained by a substitution which has the following effect. Given a two-dimensional partially directed walk in the $N-H$-plane, we insert arbitrarily long sequences of steps perpendicular to the plane containing the original walk, in positions prior to each step in the original walk, while observing self-avoidance in the three-dimensional walk thus created.  In this three-dimensional context we rename the original $N$-axis with new name, $N_1$, and create a new axis, $N_2$, perpendicular to both the $N_1$ and $H$ axes.  The projection of the resulting walk onto the $N_1-N_2$ plane is no longer self-avoiding; likewise with the $N_2-H$ plane. The projection of the resulting walk onto the $N_1-H$ plane is partially directed.  The set of valid walks in our second 3D model is a strict superset of the set of valid walks in our first 3D model.  

\begin{figure}[htbp]
\begin{center}
\includegraphics[width=0.6\textwidth]{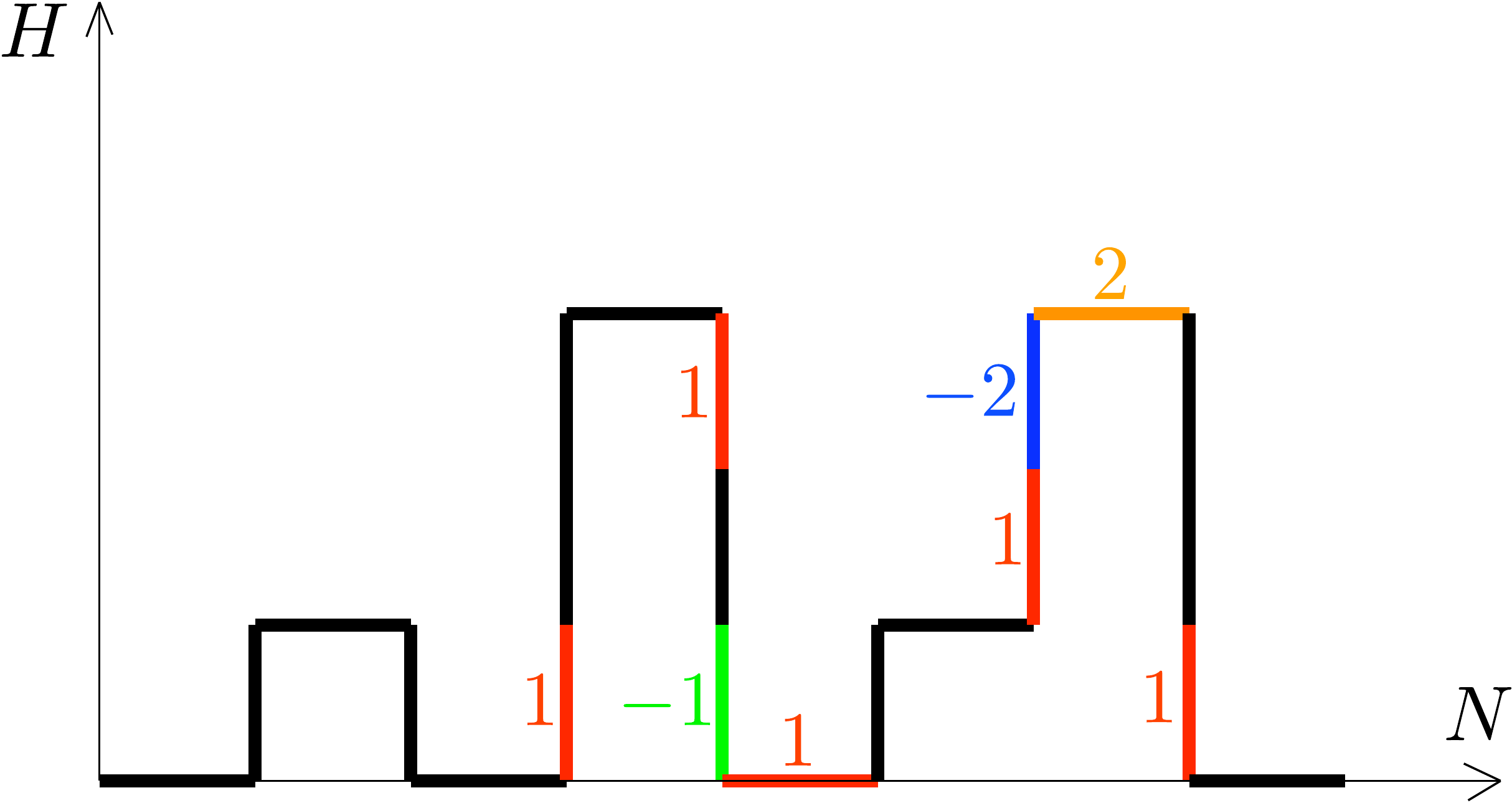}
\caption{(Color online) A $\mathbb{Z}$-colored walk with colors $-2, -1, 0, 1, 2$ occuring, where $-2$ is shown in blue, $-1$ is shown in green, $0$ is (implicitly) black, $1$ is red and $2$ is orange.}
\label{fig:2DwalkZcolored}
\end{center}
\end{figure}

\begin{figure}[htbp]
\begin{center}
\includegraphics[width=0.8\textwidth]{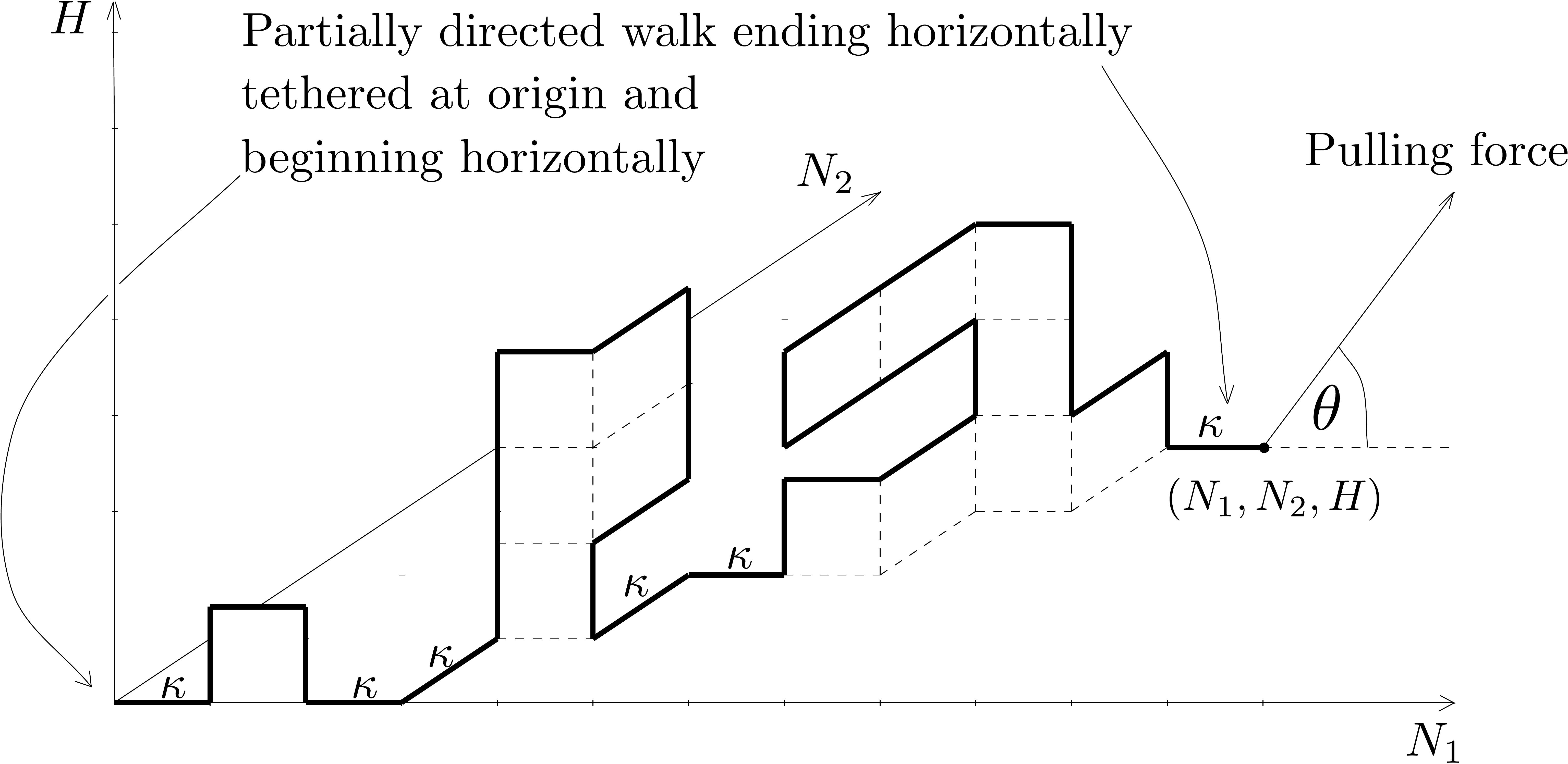}
\caption{A tethered, partially directed walk in 3 dimensions, on a sticky surface with contact weights $\kappa$ within the $N_1-N_2$ surface and a pulling-force applied at fixed angle $\theta$ within the $N_1-H$ plane.  This walk ends at position $(N_1, N_2, H) = (7,4,0)$ and has weight $\lambda^7t^{31} \kappa^5 \mu^0$.}
\label{fig:walk3DbSecondKind}
\end{center}
\end{figure}

In order to construct these three-dimensional walks  we introduce a \emph{$\mathbb{Z}$-coloring} of the two-dimensional partially directed walks that we have already defined, by associating with each edge a color drawn from the set $\mathbb{Z}$.  These \emph{$\mathbb{Z}$-colored walks} are somewhat analogous to the two-colored walks in \cite{iliev2010}, in which every horizontal edge was assigned one of two colors.  In our context we assign all possible edge-colorings to all possible walks.  Now, given a walk with a fixed coloring, we insert before an edge of color $k$, a sequence of $|k|$ edges in the positive or negative $N_2$-direction, as given by the sign of $k$.  In this way, every coloring of a two-dimensional walk produces a unique three-dimensional walk.  An example of a $\mathbb{Z}$-colored walk is given in Figure~\ref{fig:2DwalkZcolored}.  The corresponding three-dimensional walk is illustrated in Figure~\ref{fig:walk3DbSecondKind}.


The associated substitution rules for the weights can be derived as follows. In the bulk (that is, for edges whose initial vertex does not lie in the surface), a $k$-colored vertical edge of weight $y$ gets replaced by $1+|k|$ edges of combined weight $t^{1+|k|}$, and summing over all colors leads to the substitution
\be
y \gets t \left (\ldots + t^2 + t + 1 + t + t^2 + \ldots \right) = t \frac{1+t}{1-t}\;,
\ee
and a $k$-colored horizontal edge of weight $x$ gets replaced by $1+|k|$ edges of combined weight $\lambda t^{1+|k|}$, leading to
\be
x \gets \lambda t \frac{1+t}{1-t}\;.
\ee
In the surface (that is, for edges both of whose vertices lie in the surface), a $k$-colored edge of weight $\kappa x$ gets replaced by $1+|k|$ edges of combined weight $\lambda\kappa t (\kappa t)^{|k|}$, and summing over all colors leads to the substitution
\be
\kappa x \gets 
\lambda\kappa t \frac{1+\kappa t}{1- \kappa t}\;.
\ee
Finally, a $k$-colored edge that leaves the surface, having weight $\omega y$, gets replaced by $1+|k|$ edges of combined weight 
$t (\kappa t)^{|k|}$, and summing over all colors leads to the substitution
\be
\omega y\gets  
 t \frac{1+\kappa t}{1- \kappa t}\;.
\ee
The generating function for this three-dimensional model (or, equivalently, for the $\mathbb{Z}$-colored walks), is therefore given by the following substitution
\be\label{eq:subst-3d2}
(x, y, \kappa, \omega; \mu) \gets \left( \lambda t \frac{1+t}{1-t},\; t\frac{1+t}{1-t}, \; \kappa \frac{(1-t)}{(1+t)}  \frac{(1+\kappa t)}{(1 - \kappa t)}, \;
\frac{(1-t)}{(1+t)}  \frac{(1+\kappa t)}{(1 - \kappa t)}; \; \mu
 \right)\,.
\ee

\section{Singularity Analysis and Phase Diagrams}

Phase transitions in the two and three dimensional models, as path length tends to infinity, occur when singularities of the generating function closest to the origin coincide.  

We calculate the singularities of the generating function by considering the discriminant of the quadratic equation $(G-G_\text{p})(G-G_\text{np}) \equiv G^2-(G_\text{p}+G_\text{np})G+G_\text{p}G_\text{np}$ satisfied by the generating function, where $G_\text{p}$ and $G_\text{np}$ are respectively the physical and non-physical solution, corresponding to substituting roots $\mu_\text{p}$ and $\mu_\text{np}$ of Equation (\ref{eq:quadraticForMu}) into Equation~(\ref{eq:Gfull}).  

The discriminant, in original variables, is
\be \label{eq:delta_xy}
\Delta = \Delta_1 \Delta_2 
\ee
where
\be \label{eq:delta_xyFactor1}
\Delta_1 = 
\,{\frac {{x}^{2} \left( y-1 \right)  \left( y+1 \right)  \left( y+xy-x+1 \right)  \left( y+xy-1+x \right) }{ \left( -\mu+y+y{\mu}^{2}-{y}^{2}\mu-x\mu\,{y}^{2}+x\mu \right) ^{2}}}
\ee
and
\be \label{eq:delta_xyFactor2}
\Delta_2 = 
\frac
{{\kappa}^{2}{\omega}^{2} \left ( \kappa x y \mu(y-\mu-y\omega + \mu \omega) -\kappa y (1-y\mu + \mu^2)-\mu(1-\kappa-y\mu)\right )^2}
{\left (\kappa^2x^2(1 - \omega y^2)(1-\omega) + \kappa \omega y^2(1+x-\kappa x)-\kappa x (2-\omega -\kappa \omega)-\kappa \omega + 1\right )^2}\;.
\ee

\subsection{The 2D model: singularity analysis} \label{sec:2dSingularity}

In two dimensions, with path-length generating variables given by substitutions~(\ref{eq:2Dsubs}) into Equation~(\ref{eq:delta_xy}), the discriminant of the quadratic for $G$ becomes
\be \label{eq:delta2d}
\Delta = 
\frac
{\kappa^2 \lambda^2 t^2  (\mu t - 1)^2 \left (\kappa t- (\kappa-1)\mu \right)^2 (t-1)(t+1)\left (\lambda t^2 -(\lambda-1)t+1\right)\left (\lambda t^2 +(\lambda+1)t-1\right)}
{\mu (\kappa-1) \left (\lambda  t^3+ t^2 -\left(\lambda + \mu +\dfrac{1}{\mu}\right)t +1\right )^2\left (\kappa \lambda t^3 - \dfrac{\kappa}{\kappa-1} t^2 -\kappa \lambda t + 1\right )^2}
\ee

We find that the relevant singularities of $G$ in $t$ are given by roots of  
\begin{subequations}
\bea
\lambda t^2 +(\lambda+1)t-1\label{eq:term3}\\
\lambda  t^3+ t^2 -\left(\lambda + \mu +\frac{1}{\mu}\right)t +1\label{eq:term1}\\
\kappa \lambda t^3 - \frac{\kappa}{\kappa-1} t^2 -\kappa \lambda t + 1\label{eq:otherSimplePole}
\eea
\end{subequations}
The first gives rise to algebraic square-root singularities, whereas the latter two give rise to simple poles.  

When $\mu=1$ in (\ref{eq:term1}), it factorizes as 
\be
(t-1)(\lambda t^2 + (\lambda+1)t-1)
\ee
and thus contains (\ref{eq:term3}) as a factor.  Thus for $\mu=1$ the algebraic singularities coincide with poles 
given by the roots of (\ref{eq:term1}). This coincidence implies the occurrence of divergent square-root singularities (as opposed to convergent square-root singularities when $\mu\neq1$). When $\mu > 1$, the pole given by the smallest root in $t$ of (\ref{eq:term1}) dominates the algebraic singularities.

\begin{figure}[htbp]
\begin{center}
\includegraphics[width=0.6\textwidth]{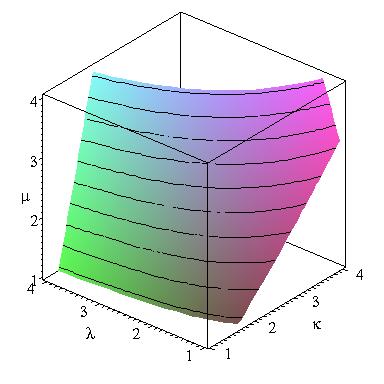}  
\caption{The surface in phase-space given by Equation~(\ref{eq:phaseTransition}), relating $\lambda$, $\mu$ and $\kappa$.}
\label{fig:phase_lmk}
\end{center}
\end{figure}

The simple poles arising from  the smallest roots in $t$ of (\ref{eq:term1}) and (\ref{eq:otherSimplePole}) coincide 
when
\be \label{eq:phaseTransition}
\lambda=\frac{\kappa\mu(\kappa-1-\kappa\mu^2)}{(\kappa-1)[(\kappa-1)^2-\kappa^2\mu^2]}\;,
\ee
which defines a surface in the space defined by $\lambda$, $\mu$ and $\kappa$, and is shown in 
 Figure~\ref{fig:phase_lmk}.  This surface separates the space into two regions and a boundary between them.  In one region, the smallest root in $t$ of (\ref{eq:term1}) is the singularity closest to the origin.  In the other region, the smallest root in $t$ of  (\ref{eq:otherSimplePole}) is the singularity closest to the origin.  On the boundary surface, both of these singularities coincide.
To interpret the meaning further, it is convenient to change to physical variables.
 
\subsection{The 2D model: physical variables} \label{sec:2d_physVar}

The relevant physical variables for our system are the temperature $T$ of the statistical mechanical ensemble of polymers, and the force $F$ by which the polymer is pulled at a fixed angle $\theta$ with respect to the horizontal axis. This leads to horizontal and vertical force components $F_x=F\cos\theta$ and $F_y=F\sin\theta$, respectively. The energy $E$ of a polymer under the influence of a pulling force is given by
\be
E = KJ - NF_x - H F_y 
\ee
where $K J $ $\equiv$ (number of contacts) $\times$ (energy per contact), $N F_x$ $\equiv$ (horizontal distance) $\times$ (horizontal force), and $H F_y$ $\equiv$ (vertical distance) $\times$ (vertical force).

The Boltzmann weight of such a configuration is then given by
\be
e^{-E/kT}=e^{-(KJ-NF_x-HF_y)/kT}\;,
\ee
which after scaling, such that the Boltzmann constant $k=1$ and the attractive energy $J=-1$, becomes
\be
e^{-E/kT}=\kappa^K\; \lambda^N\; \mu^H,
\ee
where
\be \label{eq:physVar}
\kappa = e^{\frac{1}{T}}\;,\quad\lambda = e^{\frac{F \cos \theta}{T}}\;,\quad\mu = e^{\frac{F \sin \theta}{T}}\;.
\ee

\begin{figure}[htbp]
\begin{center}
\includegraphics[width=0.6\textwidth]{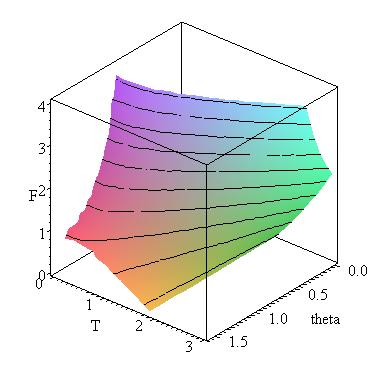}    
\caption{The 2D model: critical surface in physical variables $T$, $F$ and $\theta$. Slices through this surface at constant angle $\theta$ are shown in Figure~\ref{fig:phaseTransition_physVar-b}.}
\label{fig:phaseTransition_physVar}
\end{center}
\end{figure}

The critical surface in terms of the physical variables, whose equation is determined by substituting Equations~(\ref{eq:physVar}) into Equation~(\ref{eq:phaseTransition}), is illustrated in Figure~\ref{fig:phaseTransition_physVar}.

The surface in Figure~\ref{fig:phaseTransition_physVar} is indicative of a phase transition between an adsorbed and a desorbed state for the polymer being modeled.  This interpretation will be confirmed in Section \ref{sec:2d} where we plot the fraction of the polymer which is in contact with the surface, for various pulling angles between $0^\circ$ and $90^\circ$.    As will be seen, that part of phase space which is behind the critical surface (as pictured in Figure~\ref{fig:phaseTransition_physVar}) corresponds to a state of adsorption, in which a positive fraction of the polymer is in contact with the surface.   That part of phase space which is in front of the critical surface as pictured corresponds to a state of desorption.

\subsubsection{General observations}

\begin{figure}[htbp]
\begin{center}
\includegraphics[width=0.45\textwidth]{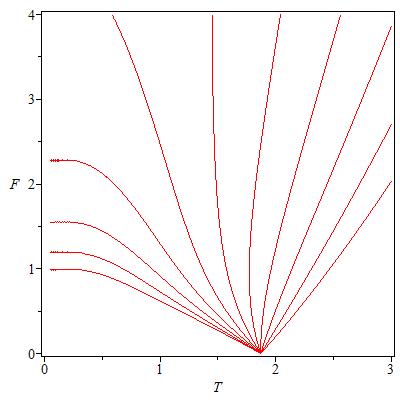}
\caption{The 2D model: temperature-force desorption transitions at different pulling angles from purely vertical to purely horizontal pulling in increments of $9^\circ$. The left-most curve corresponds to vertical pulling at $90^\circ$, and the right-most curve corresponds to horizontal pulling at $0^\circ$.}
\label{fig:phaseTransition_physVar-b}
\end{center}
\end{figure} 

Some features of the desorption transition become clearer when considering slices through the surface at constant pulling angle, leading to
temperature-force desorption curves as shown in Figure~\ref{fig:phaseTransition_physVar-b}.

When no force is applied, the polymer is adsorbed at low temperature and becomes desorbed as the temperature increases past a critical value.  This thermal desorption occurs at
\be \label{eq:critT2d}
T=1/\log(1+\sqrt2/2)\approx1.87
\ee
as can be computed by setting force equal to zero, i.e. letting $\mu=\lambda=1$ in Equation~(\ref{eq:phaseTransition}).
Naturally, in the context of zero force, there is no dependence on the angle.  When a force is applied and pulling is in a purely vertical direction, the force favors desorption, as expected.  This may be seen by tracing upwards either through the front of the phase surface as pictured in Figure~\ref{fig:phaseTransition_physVar}, or equivalently through the left-most curve within the temperature-force plots in Figure~\ref{fig:phaseTransition_physVar-b}.  (The area to the left of a curve in the temperature-force plot corresponds to a state of adsorption; and the area to the right to desorption.) As is also intuitive, pulling with a purely horizontal angle favors adsorption, since horizontally stretched polymers will favor adsorption. This may be seen by considering either the back of the phase surface as pictured, or the right-most curve in the temperature-force plot.

\subsubsection{Critical angles}

Perhaps surprisingly, we see that there is a non-zero critical angle below which pulling will never induce desorption no matter how great the force, and will eventually induce adsorption.  To determine this value, we compute the angle for which the temperature-force curve has vertical slope at $F=0$.  The force-temperature plots in Figure~\ref{fig:phaseTransition_physVar-b} show this value to be about $27^\circ$; and in fact the critical angle is
\be
\theta=\tan^{-1}(1/2)\approx26.6^\circ\;.
\ee
 At zero temperature, the transition to desorption occurs when
\be
F=\frac1{\sin\theta-\cos\theta}
\ee
Hence, once $\theta\leq 45^\circ$, 
no zero-temperature desorption occurs.

\subsection{The order parameter} \label{sec:2d}

\begin{figure}[htbp]
\begin{center}\hfill
\includegraphics[width=0.4\textwidth]{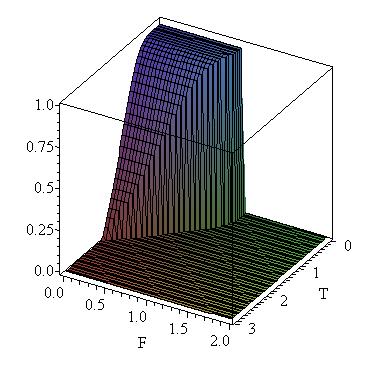}\hfill
\includegraphics[width=0.4\textwidth]{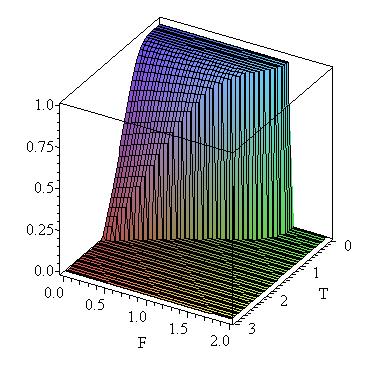}\hfill
\end{center}
\vfill
\begin{center}\hfill
\includegraphics[width=0.4\textwidth]{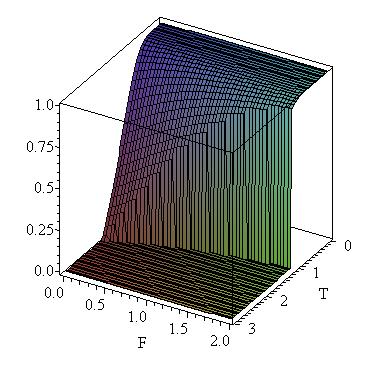}\hfill
\includegraphics[width=0.4\textwidth]{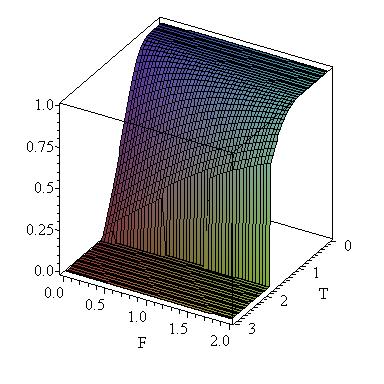}\hfill
\end{center}
\vfill
\begin{center}\hfill
\includegraphics[width=0.4\textwidth]{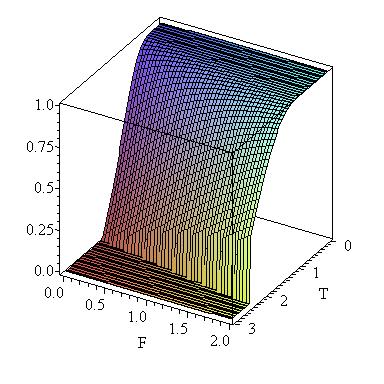}\hfill
\includegraphics[width=0.4\textwidth]{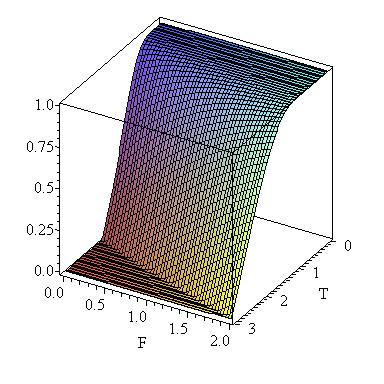}\hfill
\end{center}
\caption{The order parameter $\mathcal C$, the fraction of polymer adsorbed onto the surface, as a function of temperature $T$ and force $F$, for pulling angles $\theta=90^\circ$, $72^\circ$, $54^\circ$, $36^\circ$, $18^\circ$ and $0^\circ$ from top left to bottom right.}
\label{fig:coverage}
\end{figure} 

To confirm that there is indeed a transition between desorbed and adsorbed phases, we consider the fraction 
\be
\mathcal{C}=\lim_{L\to\infty}\frac1L\langle K\rangle
\ee
of the polymer that is in contact with the surface, which is an order parameter for the transition.
In the desorbed state, we expect $\mathcal C=0$, whereas it will be positive in the adsorbed state. For a polymer lying entirely in the surface ${\mathcal C}=1$.  A second-order phase transition between an adsorbed and a desorbed state will be reflected in a smooth change of $\mathcal C$, 
whereas a first-order phase transition will be indicated by a jump of $\mathcal C$ from zero to a non-zero value.

We calculate $\mathcal C$ from
\be \label{eq:coverage}
\mathcal{C}=\frac{\partial\log t_c}{\partial\log\kappa}\;,
\ee
i.e. from the change of the critical fugacity $t_c$ with respect to the interaction weight $\kappa$ . 
In Figure~\ref{fig:coverage}  we show the temperature-force dependence of $\mathcal C$ at various fixed
pulling angles ranging from vertical to horizontal in decrements of $18^\circ$.
For zero force, we see that $\mathcal C$ changes continuously with temperature, indicative of a second-order phase transition. As shown
in the bottom right diagram of Figure~\ref{fig:coverage}, the transition remains second-order when a horizontal pulling force is applied, and
only the location of the transition shifts.
On the other hand, as shown by all the other diagrams of Figure~\ref{fig:coverage}, the transition becomes first-order as soon as there is a
non-zero vertical component of the pulling force.
Note that once adsorbed, the value of $\mathcal C$ is independent of the pulling force $F$ and angle $\theta$.

Thus, Figure~\ref{fig:coverage} confirms that there is indeed a desorption phase transition, as indicated above.

\subsection{The first 3D model: physical variables} \label{sec:3d}

According to the substitution (\ref{eq:subst-3d1}), we now have variables $\lambda_1$ conjugate to horizontal variable $N_1$ and $\lambda_2$ conjugate to the other horizontal variable $N_2$. The singularity analysis that was carried out for two dimensions in Section~\ref{sec:2dSingularity} is nearly unchanged, since it amounts to replacing $\lambda$ with $\lambda_1+\lambda_2$ in Equation~(\ref{eq:delta2d}).

The physical variables for three dimensions are now given by substitutions 
\be
\kappa = e^{\frac{1}{T}}\;,\quad\lambda_1 = e^{\frac{F \cos \theta\cos\phi}{T}}\;,\quad\lambda_2 = e^{\frac{F \cos \theta\sin\phi}{T}}\;,\quad\mu = e^{\frac{F \sin \theta}{T}}
\ee
where, as before, $T$ is temperature, $F$ is force and $\theta$ is pulling angle upwards away from the base surface.  The extra variable $\phi$ is the component of pulling-angle measured away from the $N_1$-axis in the $N_1N_2$-plane.

\begin{figure}[htbp]
\begin{center}
\includegraphics[width=0.45\textwidth]{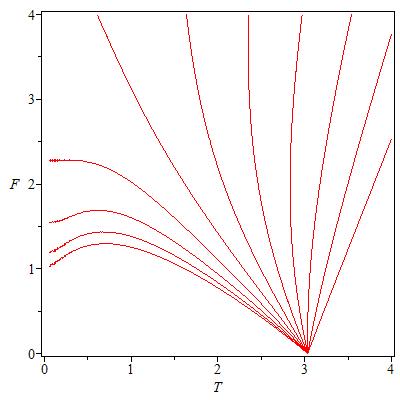}
\includegraphics[width=.45\textwidth]{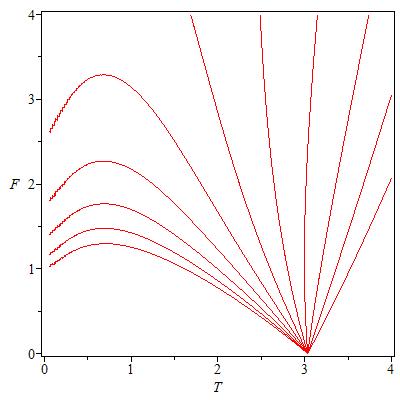}
\caption{The first 3D model: shown are temperature-force desorption transitions for horizontal pulling angle $\phi=0^\circ$ (left) and $\phi=45^\circ$ (right). In each diagram the curves correspond to different vertical pulling angles $\theta$ from purely vertical to purely horizontal pulling, in increments of $9^\circ$. The left-most curve corresponds to vertical pulling at $90^\circ$, and the right-most curve corresponds to horizontal pulling at $0^\circ$.}
\label{fig:phaseTransition_physVar3d}
\end{center}
\end{figure} 

As for the two-dimensional case discussed above in Subsection~\ref{sec:2d_physVar}, we consider temperature-force desorption curves at constant vertical pulling angles $\theta$. In Figure~\ref{fig:phaseTransition_physVar3d} we show the resulting diagrams for two different horizontal pulling
angles $\phi=0^\circ$ and $\phi=45^\circ$. The pictures are qualitatively similar to Figure~\ref{fig:phaseTransition_physVar} for two dimensions, 
with a few interesting differences.

The phenomenon of \emph{reentrance} is observable in both diagrams of Figure~\ref{fig:phaseTransition_physVar3d}, as is seen by the shape of the temperature-force curves for steep angles. As may be observed in either diagram of the figure, when pulling vertically with a force $F$ slightly larger than one, the polymer is desorbed at sufficiently high temperature, gets adsorbed upon decreasing the temperature, but then desorbs again when the temperature is decreased even further towards zero.

Notably, there is a difference between the two diagrams, in that in the left diagram ($\phi=0^\circ$) the reentrance phenomenon becomes weaker and disappears upon decreasing $\theta$, whereas in the right diagram ($\phi=45^\circ$) the reentrance phenomenon persists
up until a critical value of $\theta$, at which zero-temperature desorption disappears completely, is reached. 

Reentrance has been observed before in related models, see e.g.  \cite{krawczyk2004}.  An explanation of the occurrence of reentrance 
is provided by a zero-temperature entropy argument, which shows that the critical force changes for small temperature $T$ linearly as
\be
F\approx 1+TS
\ee
where $S$ is the the configurational entropy available to the walks adsorbed onto the two-dimensional surface.  

In this three-dimensional model walks adsorbed onto the surface are fully directed. When pulling at a horizontal angle $\phi=45^\circ$, all  
$2^L$ possible adsorbed configurations of walks of length $L$ are equally likely, and hence the configurational entropy $\log 2$ is positive. 
This effect is clearly visible in the temperature-force diagram via the positive slope of the curves for low temperature.

When pulling at a horizontal angle $\phi\neq45^\circ$, there is a preferred direction for the stretched walk, hence the configurational entropy is zero. This effect is visible in the temperature-force diagram for $\phi=0$, where a non-vertical pulling angle $\theta$ immediately leads to
a horizontal slope of the curves for low temperature. We argue that this effect is ultimately responsible for the disappearance of the reentrance phenomenon.

As for the two-dimensional model, we can calculate several quantities exactly. For example,  the critical value of thermal desorption  
is given by
\be \label{eq:critT3d}
T=1/\log(7/8+\sqrt{17}/8)\approx3.03\;.
\ee
For $\phi=45^\circ$ the non-zero critical angle below which pulling cannot induce desorption and will eventually induce 
adsorption is given by
\be
\theta=\tan^{-1}((1+\sqrt{17}/17)\sqrt2/4)\approx24^\circ\;,
\ee
zero-temperature desorption occurs when
\be
F=\frac2{2\sin\theta-\sqrt2\cos\theta}\;,
\ee
and the critical angle below which no zero-temperature desorption occurs is
\be
\theta = \tan^{-1}(\sqrt2/2)\approx35^\circ\;.
\ee

\subsection{The second 3D model: singularity structure and physical variables} \label{sec:3e}

After the substitution (\ref{eq:subst-3d2}), we obtain variables as in the two-dimensional model, i.e. we only vary the vertical pulling angle
$\theta$ (analogous to letting $\phi=0$ in the first 3D model). The singularity analysis that was carried out for two dimensions in Section~\ref{sec:2dSingularity} needs to be done carefully, 
as the substitution (\ref{eq:subst-3d2}) potentially changes the phase diagram. The resulting algebraic equations become rather large
and cumbersome, for example the equation corresponding to the surface given by Equation~(\ref{eq:phaseTransition}) for the two-dimensional model now
becomes an algebraic equation in $\lambda$, $\kappa$, and  $\mu$ which involves 932 monomials. We are therefore restricted to 
performing a numerical analysis.

The physical variables are
\be
\kappa = e^{\frac{1}{T}}\;,\quad\lambda= e^{\frac{F \cos \theta}{T}}\;,\quad\mu = e^{\frac{F \sin \theta}{T}}
\ee
where, as before, $T$ is temperature, $F$ is force and $\theta$ is pulling angle upwards away from the base surface.

\begin{figure}[htbp]
\begin{center}
\includegraphics[width=0.45\textwidth]{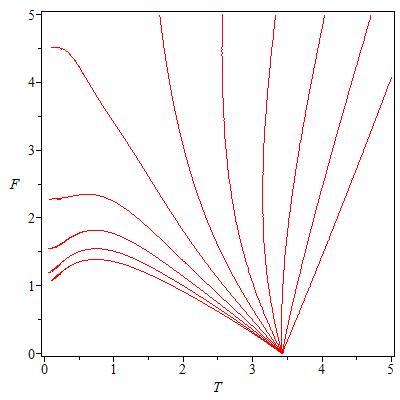}
\caption{The second 3D model: shown are temperature-force desorption transitions for different vertical pulling angles $\theta$ from purely vertical to purely horizontal pulling, in increments of $9^\circ$. The left-most curve corresponds to vertical pulling at $90^\circ$, and the right-most curve corresponds to horizontal pulling at $0^\circ$.}
\label{fig:phaseTransition_physVar3d-2}
\end{center}
\end{figure} 

We again consider temperature-force desorption curves at constant pulling angles $\theta$ away from the $N_1-N_2$ plane. In Figure~\ref{fig:phaseTransition_physVar3d-2} we show the resulting diagram. The picture is qualitatively very similar to the diagram 
for pulling at $\phi=0^\circ$ in Figure~\ref{fig:phaseTransition_physVar3d} for the first three-dimensional model.

The phenomenon of reentrance is again observable in Figure~\ref{fig:phaseTransition_physVar3d}. Similarly to what was observed in the
 temperature-force diagram of the first three-dimensional model for $\phi=0$,  a non-vertical pulling angle $\theta$ immediately leads to
a horizontal slope of the curves for low temperature and ultimately to the disappearance of the reentrance phenomenon for sufficiently
shallow pulling. We argue that this is again due to the fact that, with pulling as described, a stretched polymer adsorbed onto the surface at $T=0$ has zero configurational entropy.

\section{Conclusion}

Na\"{\i}vely, one might expect that pulling sufficiently hard on a polymer tethered to a horizontal surface, at any angle containing a non-zero vertical component away from that surface, would eventually affect desorption.  We have shown that in our models this is not the case, and that pulling with any positive angle below a certain critical value will induce adsorption. 
With our study of three-dimensional models, we have also shown how the reentrance phenomenon, which is present in force-induced
polymer desorption, is affected by changing the entropy of the adsorbed polymer due to pulling with a horizontal force component. Interestingly, in semi-flexible polymers, which are modeled on a lattice by weighting bends, hence making the
polymer stiffer, reentrance is affected in a somewhat similar manner \cite{iliev2010}. It would be interesting to extend our study by including the effect of stiffness.

\section*{Acknowledgements}
Financial support from the Australian Research Council via its support
for the Centre of Excellence for Mathematics and Statistics of Complex
Systems is gratefully acknowledged by the authors. J Osborn thanks the
School of Mathematical Sciences, Queen Mary, University of London as well as the Computing Laboratory, University of Oxford, for
hospitality, and T Prellberg thanks the Centre for Mathematics and its Applications, The Australian National University, for hospitality.


\begin{thebibliography}{10}

\bibitem{Rensburg2000} E. J. Janse van Rensburg, {\em The Statistical Mechanics of Interacting Walks, Polygons, Animals and Vesicles},
Oxford University Press, Oxford, 2000. 

\bibitem{Rensburg2003} E. J. Janse van Rensburg, J. Phys. A. {\bf 36}, R11 (2003).

\bibitem{kumar2010} S. Kumar and M. S. Li, Physics Reports {\bf486},  1 (2010). 

\bibitem{mishra2003} P. K. Mishra, S. Kumar, and Y. Singh, Physica A, {\bf 323} 453 (2003). 

\bibitem{mishra2005} P. K. Mishra, S. Kumar, and Y. Singh, Europhys. Lett. {\bf 69} 102 (2005). 

\bibitem{orlandini2004} E. Orlandini, M. C. Tesi, and S. G. Whittington, J. Phys. A. {\bf 37}, 1535 (2004).

\bibitem{owczarek2009} A. L. Owczarek, J. Stat. Mech., P11002 (2009). 



\bibitem{owczarek2010} A. L. Owczarek, J. Phys. A. {\bf 43}, 225002 (2010). 

\bibitem{iliev2010} G. K. Iliev, E. Orlandini, and S. G. Whittington, J. Phys. A. {\bf 43}, 315202 (2010). 

\bibitem{bhattacharya2009} S. Bhattacharya, V. G. Rostiashvili, A. Milchev, and T. A. Vilgis, Macromolecules, {\bf 42}, 2236 (2009). 

\bibitem{krawczyk2004} J. Krawczyk, T. Prellberg, A. L. Owczarek, and A. Rechnitzer, J. Stat. Mech., P10004 (2004). 

\bibitem{kumar2007} S. Kumar and D. Giri, Phys. Rev. Lett. {\bf  98}, 048101 (2007)

\bibitem{rajesh2008} R. Rajesh, D. Giri, I. Jensen, and S. Kumar, Phys. Rev. E {\bf 78}, 021905 (2008)

\bibitem{jensen2010} I. Jensen, D. Giri, and S. Kumar, Mod. Phys. Lett. B {\bf 24}, 379 (2010)

\bibitem{prodinger2004} H. Prodinger, S\'em. Lothar. Combin. {\bf 50}, B50f (2004). 





\end{thebibliography}
\end{document}